# Lasing action in active dielectric nanoantenna arrays

*Son Tung Ha[†], Yuan Hsing Fu[†], Naresh Kumar Emani[†§], Zhenying Pan, Reuben M. Bakker, Ramón Paniagua-Domínguez, and Arseniy I. Kuznetsov*[*]

Data Storage Institute, A*STAR (Agency for Science, Technology and Research), 138634, Singapore

[*]Corresponding author, email: arseniy_k@dsi.a-star.edu.sg

[†]These authors equally contributed to this work.

[§]Present affiliation: Indian Institute of Technology, Hyderabad, India

**Abstract**

Directional lasing, with a low threshold and high quality factor, in active dielectric nanoantenna arrays is demonstrated. This is achieved through a leaky resonance excited in coupled gallium arsenide (GaAs) nanopillars. The leaky resonance is formed by partially breaking a bound state in the continuum (BIC) generated by the collective, vertical electric dipole resonances excited in the nanopillars for sub-diffractive arrays. By opening an unprotected, diffractive channel along one of the periods of the array one can control the directionality of the emitted light without sacrificing the high Q associated with the BIC mode, thus achieving directional lasing. A quality factor Q = 2750 is achieved at a controlled angle of emission of ~ 3° with respect to the normal of the array with a pumping fluence as low as 10 μJ/cm². We demonstrate the possibility to control the lasing directivity and wavelength by changing the geometrical parameters of the nanoantenna array, and by tuning the gain spectrum of GaAs with temperature. Lasing action is demonstrated at different wavelengths and emission at different angles, which can be as large as 25° to the normal. The obtained results provide guidelines for achieving surface emitting laser devices based on active dielectric nanoantennas that are compact and highly transparent.

**Keywords:** dielectric nanoantennas, laser, nanolaser, bound state in the continuum, Mie resonance, gallium arsenide

High-index dielectric nanoparticles supporting strong electric and magnetic resonances have drawn increasing attention in recent years.[1-4] Their low dissipative losses at visible and near IR wavelengths provide a more promising way to efficiently manipulate light at nanoscale as compared to conventional plasmonics.[5,6] Another advantage of many suitable dielectric materials is their compatibility with semiconductor fabrication processes which is promising for on-chip integration in photonic devices. Finally, the wealth of optical modes accessible in dielectric nanoparticles even in their simplest shapes, and their associated interference effects, bring exciting opportunities to realize optical antennas with strong directionality that are straight forward to fabricate. Due to these advantages, many applications of resonant dielectric nanostructures have been proposed and demonstrated in the last few years, such as: beam bending and switching,[7-13] flat lenses,[14-17] nonlinear harmonic generation,[18-24] directional scattering,[25-27] hologram and vortex beam generation,[8,9,28-32] and polarization control.[31-33]

A particularly important characteristic of any photonic system is an ability to produce amplified spontaneous or stimulated emission. While dielectric resonant nanostructures have been studied for light confinement and enhanced spectroscopies (Raman and fluorescence[34-41]), up to now, despite some theoretical proposals[42], there has been no experimental demonstration of lasing action based on those. The main reasons can be attributed to the low quality factor (Q) of the resonant modes and, thus, the lack of suitable cavity designs.

Recently, a novel way of light confinement has been demonstrated by means of so called bound states in the continuum (BIC). In these states, light is localized within the structure supporting the BIC mode, despite the mode lying in the continuum part of the spectrum, therefore coexisting with the radiative modes[43-46]. Theoretically, the Q factor of these modes can reach infinity if the BIC condition is strictly satisfied[45]. In practice, the Q factor is limited by the finite size of real systems, which implies the existence of unprotected channels and renders the BIC into so called supercavities or leaky resonances[45,47-49] (the latter may emerge even for infinite systems[49]). Nevertheless, the high Q values achievable even in real, practical, systems supporting these modes, make them perfectly suitable for lasing, and some devices working under this principle have recently been demonstrated.[50-56] A common limitation of these devices is the poor directionality of the emitted light, which is radiated into free space by defects or simply by the edges of the device.[57]

Here we propose a novel design that takes advantage of both dielectric nanoantenna resonances and BIC confinement to experimentally realize lasing in a 2-dimensional (2D) array of resonant dielectric nanostructures with controlled directionality.

The BIC is formed at the gamma ($\Gamma$) point of the array, *i.e.* for $\mathbf{k}_{||} = (k_x, k_y) = \mathbf{0}$, when the particles support only resonant vertical dipole mode (*i.e.* oriented along the z-axis, normal to the array) and the array is strictly sub-diffractive and of infinite size in x and y directions (see Fig.1a&b). In this situation it is easy to visualize how the BIC is formed. For a sub-diffractive, 2D array of dipoles oscillating in phase, radiation is only allowed in the normal direction to the array (in all other directions, destructive interference from the rest of the dipoles in the array leads to zero net radiation). In the case in which these dipoles oscillate precisely in the normal direction, and therefore do not radiate towards it, no radiation from the system is allowed, and the BIC is formed. This BIC is topologically robust. The polarization vector forms a vortex around the $k_x = k_y = 0$ point with topological charge $q = 1$ (and nodal lines $E_x = 0$ along $k_x = 0$ and $E_y = 0$ along $k_y = 0$), and therefore cannot be removed unless strong variations in the geometrical parameters of the system are introduced.[58] Now, by carefully adjusting one of the periods to support a diffraction order at the resonant wavelength, a leaky channel can be opened.[49] In this way, it is possible to control the directionality of the emission while still maintaining a very high confinement factor of the cavity near the vertical direction. This design can be applied to various active high-index semiconductor materials to achieve lasing at visible and IR wavelengths. Moreover, the low surface footprint of our laser structure leads to high transparency in the wavelength range of interest (*e.g.,* >85% over 700 - 900 nm). This, to the best of our knowledge, is the first surface emitting laser with such high transparency, which might have significant applications in multi-layered photonic devices.

Our laser device consists of vertical GaAs nanopillars as building blocks with a diameter of ~ 100 nm and a height of 250 nm. GaAs is a widely used, III-V semiconductor with a near IR direct bandgap at ~1.42 eV(300K) and a relatively high refractive index of ~ 3.5.[59,60] Recently, GaAs-based metasurfaces have also been studied for second harmonic generation and optical modulation.[20,21,61-63] In our device, both vertical and in-plane dipole resonances are supported in the nanopillars as illustrated in **Figure 1a**. The nanopillars are arranged in a 2-dimensional (2D) rectangular array, supported by a quartz substrate and embedded in silica (SiO$_2$), as depicted in **Figure 1b**. In one of the directions, the period of the lattice is sub-diffractive for wavelengths within the photoluminescence (PL) band of GaAs (namely *x*-axis, $P_x$ = 300 nm) while in the other direction, the period is diffractive (namely *y*-axis, $P_y$ = 540 nm). This design opens up a radiation channel that turns the BIC mode of the purely sub-diffractive case into a leaky resonance with finite Q (see more details in Supplementary Information Figure S1 and S2) and presents a cross-point with $k_y = 0$ in the emission plane, as schematically depicted in Figure 1b.

The fabrication process of the device starts with *i*-GaAs (500 nm)/AlAs(100 nm)/*i*-GaAs (625 µm) wafers (Semiconductor Wafer Inc). The AlAs layer is dissolved in a dilute HF solution(~5 wt. % in $H_2O$), and the GaAs top film is transferred to fused silica substrate using an epitaxial lift-off procedure.[64,65] The GaAs film is etched to the design thickness of ~ 250 nm by dry etching using $Cl_2$ chemistry (ICP-RIE, Oxford Plasmalab 100). The arrays are written in a negative e-beam resist, hydrogen silsesquioxane (HSQ), using e-beam lithography. After developing the resist, the GaAs pillar patterns are realized by another RIE step using the same etching chemistry as above. The detailed fabrication process is described in Supplementary Information Figure S3. The scanning electron microscope (SEM) images of one of the fabricated GaAs arrays are shown in **Figure 1c**. The GaAs arrays are then covered in a thick layer of HSQ resist by repeated spin-coating and thermal curing steps. The curing process cross links and hardens the HSQ resist resulting in the formation of Spin On Glass overcoat with refractive index close to 1.5.

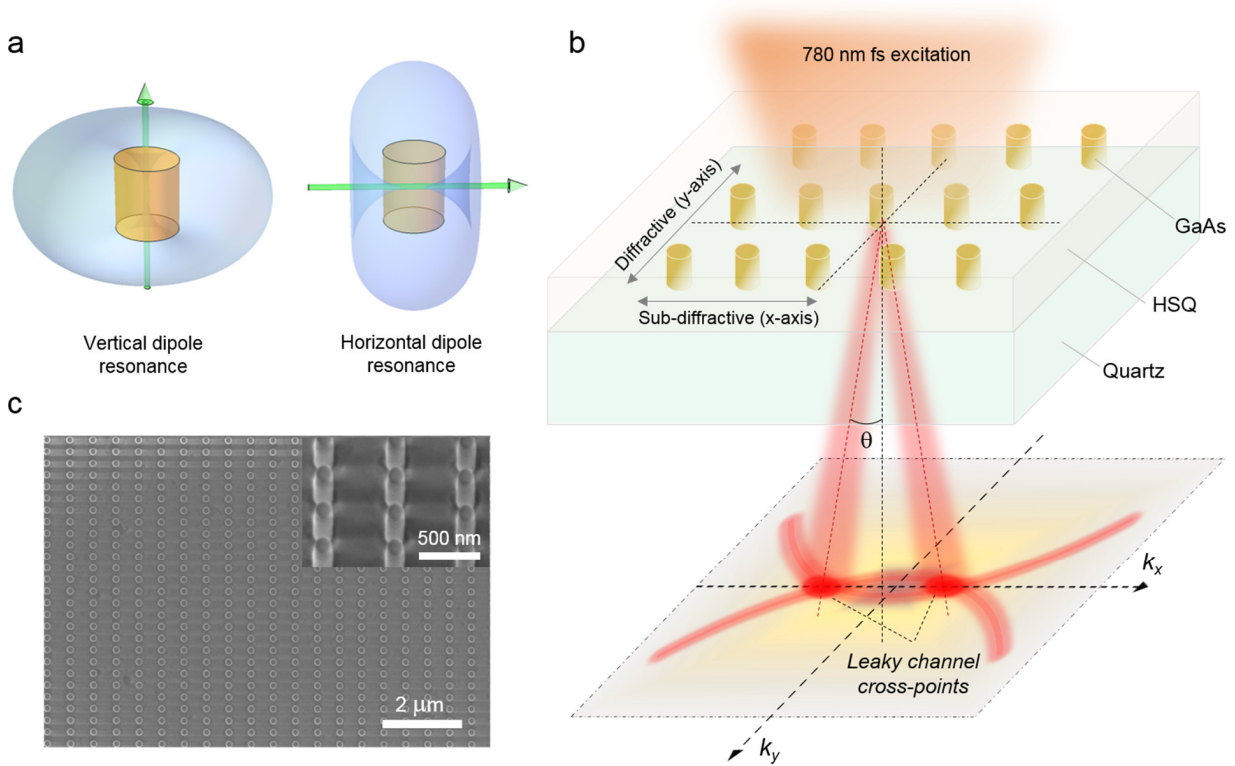

**Figure 1. Structure of the resonant dielectric nanoantenna array. a**, Resonant modes in a dielectric nanopillar showing the vertical and horizontal dipole resonances. The GaAs pillar in our laser device has a diameter of 100 nm and a height of 250 nm. **b**, Schematics of the GaAs nanopillar array on a fused silica substrate embedded in Spin On Glass. Along the y-axis, the period ($P_y$ = 540 nm) is designed to support diffraction at the emission wavelength range of GaAs (~ 830 nm at 77K). Along the x-axis, the period is fixed at 300 nm and is sub-diffractive. **c**, SEM image of the fabricated array at normal incidence. Inset: SEM image at 30° tilted angle showing the cylinder shape.

To investigate the resonant modes in the GaAs array, we performed spectrally-resolved back focal plane imaging[17] using a 50X objective with a numerical aperture (NA) of 0.55. The measurements are performed with an inverted optical microscope setup (Nikon Ti-U) coupled to a spectrometer (Andor SR-303i) equipped with an EMCCD detector (Andor, Newton)[5]. The schematic of the measurement setup is illustrated in **Figure 2a**. Light, from a halogen lamp, polarized along the x-axis is focused on the sample surface via a top objective (50X, 0.55 NA). The transmission signal is then collected by a lower objective with the same specifications. The back focal plane of the bottom objective is imaged onto the entrance slit of the spectrometer. The slit has a width of 50 µm and is oriented along the x-axis of the sample. This measurement provides angular-resolved transmission spectrum of the sample in a single image (see Fig.2b). Angular information is recorded along the slit axis of the spectrometer (*i.e.,* corresponding to a variation of the angle of incidence contained in the xz-plane of the sample). The maximum detected angle is determined by the NA of the objective and is ~ 33 degrees for the lens used.

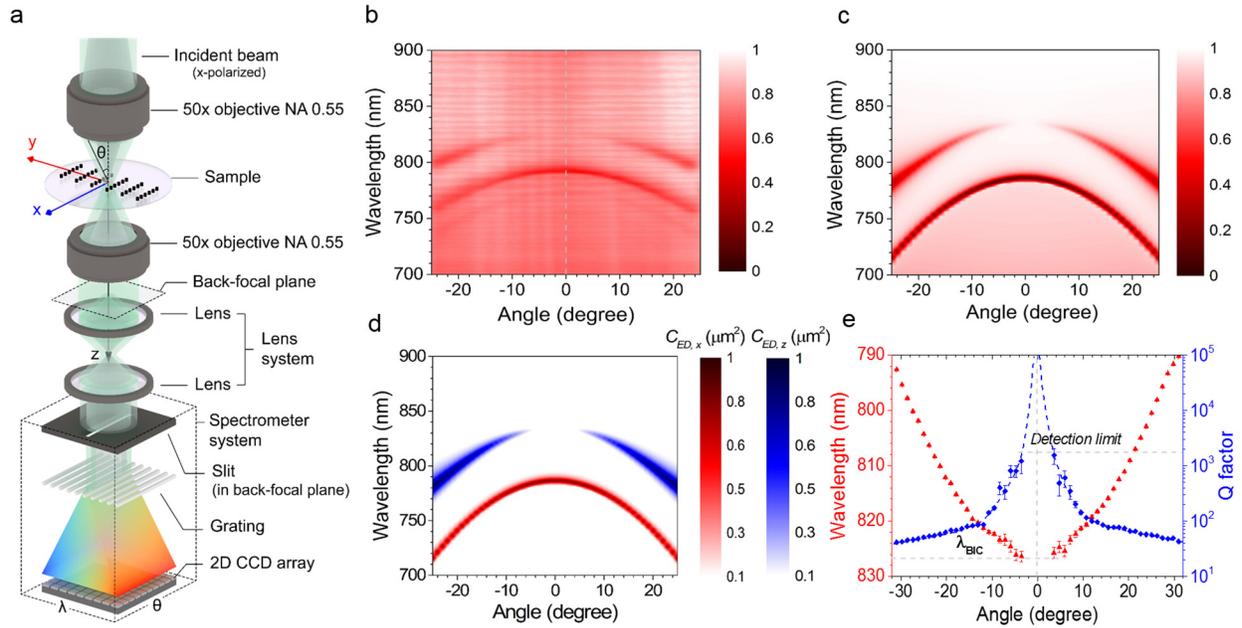

**Figure 2. Resonant modes in the GaAs nanopillar array. a**, Schematic of the spectrally-resolved back focal plane imaging measurements. **b**, Angular-resolved transmission spectrum image measured for an array consisting of GaAs nanopillars with 100 nm diameter, 250 nm height and 540 nm period along y-axis. The period along x-axis is 300 nm. The measured data correspond to a varying angle of incidence contained in the xz-plane of the sample and p-polarization. **c**, Simulated transmission as a function of angle and wavelength of incidence. **d**, Numerically calculated multipolar decomposition, showing the two modes excited in the nanopillars as a function of angle and wavelength of incidence, namely a horizontal electric dipole (red color scale) and a vertical electric dipole (blue color scale). The simulations show that the resonance dip observed in transmission at ~830 nm for oblique incidence corresponds to the excitation of a vertical electric dipole resonance, which vanishes at 0° forming the BIC. **e**, Angular dependence of wavelength and Q factor for the vertical electric dipole mode extracted from b showing the emergence of the BIC with Q factor growing to infinity at $\lambda_{BIC}$ ~ 830 nm . The resonance wavelength shifts to higher energy and the Q factor decreases when the angle increases.

**Figure 2b** shows the measured transmission for p-polarized light (polarization along x axis at normal incidence) passing through the GaAs nanopillar array as function of the wavelength and the angle of incidence. The results show two resonant bands in the system (manifested as dips in transmission), the lowest energy band narrows and then vanishes when incidence becomes normal. The transmission spectra at -9 and 0 degree, as extracted from **Figure 2b** are shown in Supplementary Information, Figure S4. As can be seen, the resonance at ~ 825 nm vanishes at 0 degree while the other mode at ~ 790 nm is red-shifted for decreasing angle of incidence. To understand the nature of these resonant modes, we performed numerical simulations transmission spectrum dependence on angular incidence using a commercial electromagnetic solver based on the Finite Element Method (Comsol Multiphysics). The results, presented in **Figure 2c**, are in a very good agreement with the experimental ones, showing two resonance modes at ~790 nm and ~825 nm for close-to-normal incidence, that blue-shift as the angle increases. Multipole decomposition of the displacement currents induced in the nanopillars is used to identify the nature of these resonances. The results of this analysis are plotted in **Figure 2d**. They clearly show that the resonant dips observed in the transmission spectra correspond to a diffractively coupled, x-oriented, in-plane electric dipole and a z-oriented, vertical electric dipole, for the high energy and low energy resonances, respectively. While not shown here, the rest of multipoles are negligible in this frequency range for this incidence condition. Details of both the numerical simulations and the multipole decomposition can be found in the Supplementary Information. The resonance at 825 nm is associated with the vertical dipole vanishes at normal incidence, corresponding to the emergence of the symmetry-protected, leaky resonance associated with the bound state at the $\Gamma$ point of the first Brillouin zone for the purely sub-diffractive array. **Figure 2e** shows the angle-dependent wavelength and Q factor of the vertical electric dipole resonance extracted from angle-resolved transmission data in Figure 2b. As can be seen, the Q factor sharply increases as the angle decreases below 10 degrees. The Q factor reaches ~ 1500 at around 4 degrees which is at the resolution limit of the spectrometer (~ 0.6 nm for 150 gr/mm grating). This behavior confirms the BIC characteristic of the resonant mode.

To demonstrate lasing, the 2D GaAs nanopillar array is optically pumped using a femtosecond laser (780 nm, 200 fs pulse width at a repetition rate of 100 KHz). The pumping laser is focused on the sample using a 5X microscope objective resulting in a laser spot diameter of ~ 40 μm. The emission signal from the array is collected using a 50X, long working distance microscope objective (NA = 0.45) and dispersed in the spectrometer using a 1200 gr/mm grating with 1000 nm blaze giving a spectral resolution of 0.1 nm. A long pass filter (800 nm, Thorlabs) is used to filter out the excitation laser line. When using GaAs, it should be noted that it presents a high surface charge recombination due to defects at room temperature, which becomes especially important in nano-scaled samples.[66,67] This leads to a poor external quantum efficiency of the material photoluminescence. In our sample, neither a surface passivation layer (*e.g.*, AlGaAs) nor a chemical treatment process was applied. To increase the emission yield, we performed the measurements at low temperature (*i.e.*, 77 - 200K), and thus the emission wavelength of GaAs is blue-shifted to 830-850 nm from the usual ~870 nm peak at room temperature.

**Figure 3a** shows the evolution of the emission spectra of the nanopillar array when pumping it at different fluences. When the pumping fluence is above 10 μJ/cm$^2$, a high-intensity, narrow peak appears at ~ 825 nm. The full width at half maximum (FWHM) of the photoluminescence also decreases from 3 nm to a minimum of 0.3 nm corresponding to a Q factor of 2750, as shown in the left inset to Figure 3a. The right inset in Figure 3a shows the output intensity in log-log scale and the FWHM of the photoluminescence as a function of the pumping fluence. The observed S-shape represents the spontaneous emission to amplified spontaneous emission to stimulated emission transition with a threshold of ~ 10 μJ/cm$^2$. **Figure 3b** shows the optical image (top) and the photoluminescence images below (middle) and above (bottom) lasing threshold. The red dashed oval in these images corresponds to the array size with the x- and y-axis labeled. The green dashed circle in the middle image represents the pumping laser spot which is about 40 um in diameter. To determine the lasing directivity, the back focal plane images of the emission is collected using a long working distance 50X microscope objective (NA = 0.45). The images and extracted directivity of the emission below and above lasing threshold are shown in **Figures 3c** and **3d**, respectively.

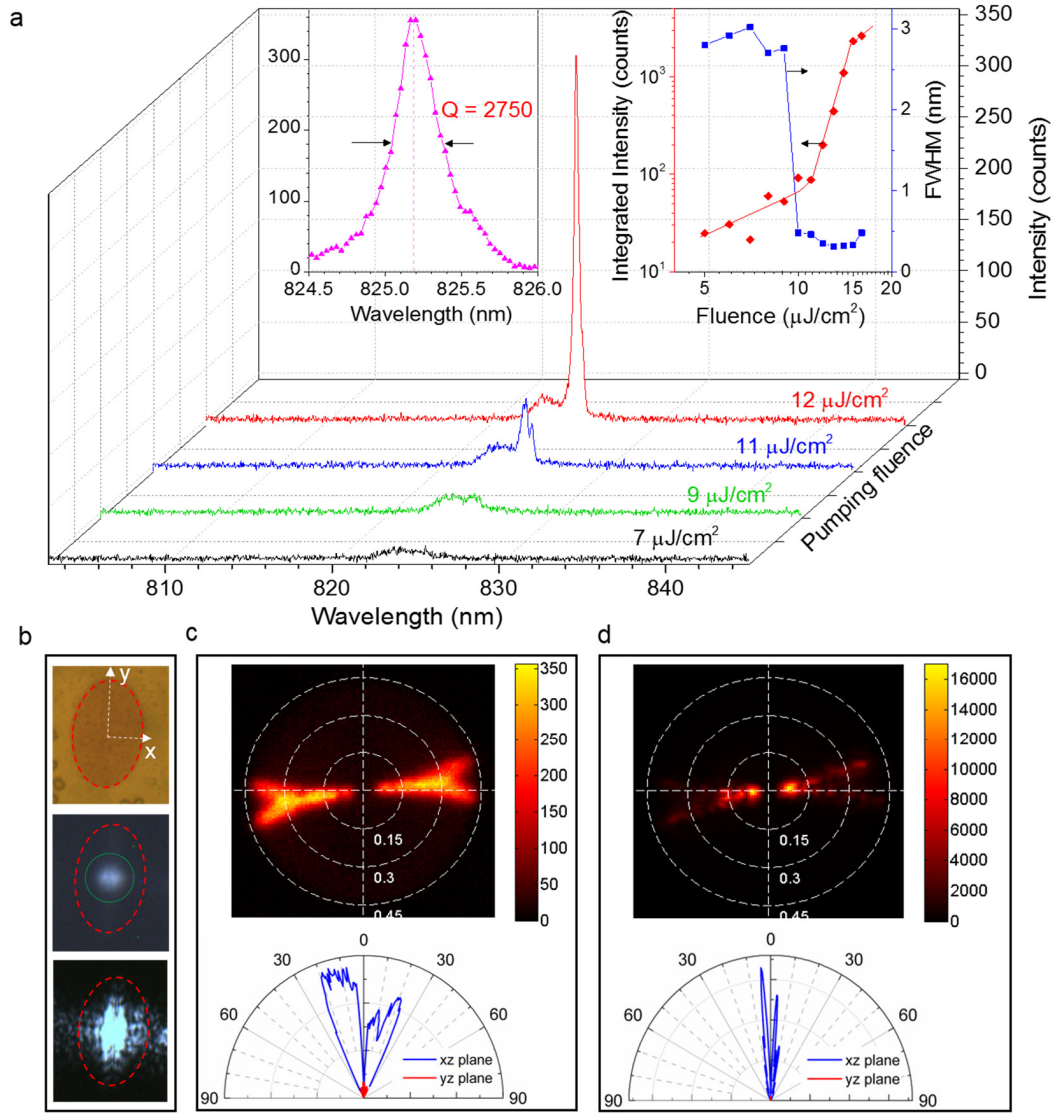

**Figure 3. Lasing action in the resonant dielectric nanoantenna array. a**, Evolution of the emission spectra of the GaAs nanopillar array at different pumping fluences. Left inset: Zoomed-in lasing peak at 13 µJ/cm² pumping fluence indicating a Q factor of 2750. Right inset: Integrated output intensity as a function of input fluence in log-log scale and full width at half maximum (FWHM) of the emission peak. The observed S-curve (red line) in the input-output dependence shows the transition from spontaneous to amplified

spontaneous and then to stimulated emission. **b**, Top panel: optical image taken under white light illumination showing the edge of the GaAs array (red-dash oval) and defined x and y axes along the rectangular lattice. Middle and bottom panels show fluorescence images under femtosecond laser excitation below and above lasing threshold, respectively. The fluorescence image is collected using a CCD camera with 820 nm band-pass filter (FWHM = 20 nm) to cut out the excitation laser line. The exposure time in the middle image is 20 times higher than that of the bottom one to clearly show the spontaneous emission. **c-d**, Back focal plane image (top) and directivity (bottom) of the emission below (c) and above (d) lasing threshold, respectively, collected by long working distance 50X microscope objective (NA = 0.45). The directivity plots were obtained as cuts along x and y axes from the back focal plane images.

As can be seen from the figure, below the threshold, the emission directivity follows two photonic bands due to the grating effect of the diffraction periodicity along the y axis. On the other hand, above the threshold, the lasing emission only happens at a specific angle (*i.e.* ~ 3 degree to normal axis) along the x-axis. This angle corresponds to the crossing point of the emission bands with $k_x$ in the emission plane (leaky channel cross points). Polarization dependence measurement of the emission intensity also shows maximum along x-axis (as shown in the Supplementary Information, Figure S5).

In general, the lasing threshold is inversely proportional to the optical gain (*g*) and the confinement factor ($\Gamma_E$): $P_{th} \sim (g(\lambda) \times \Gamma_E)^{-1}$. As discussed in Figure 2, as the angle of incidence (or, by reciprocity, the angle of emission), $\theta$, approaches 0, the resonance wavelength is red-shifted and the Q factor dramatically increases ($\Gamma_E$ increases). Thus, the lasing happens at the wavelength and angle where the minimum lasing threshold is achieved. It should be noted that the emission bands below lasing threshold, shown in Figure 3c, vanish near the normal axis. This is due to the fact that the emission is a combination of the lattice effect (emission along the diffraction directions) and the emission of vertical dipoles, which resonantly enhance the PL and which have vanishing emission at angles close to the normal direction (the oscillation direction of the dipoles). The resonantly enhanced photoluminescence is discussed in more detail in Supplementary Information Figure S6. Another interesting feature of the system is that, when higher pumping fluence is used, higher energy modes are also excited, with lower Q and higher emission angle (Supplementary Information, Figure S7 and S8). By changing the temperature of the device, it is possible to spectrally move the maximum of the gain, in such a way that the minimum threshold happens at higher angles, despite having a lower Q factor. In this way, it is possible to obtain directional lasing at an angle as large as ~ 25 degree (Supplementary Information, Figure S9).

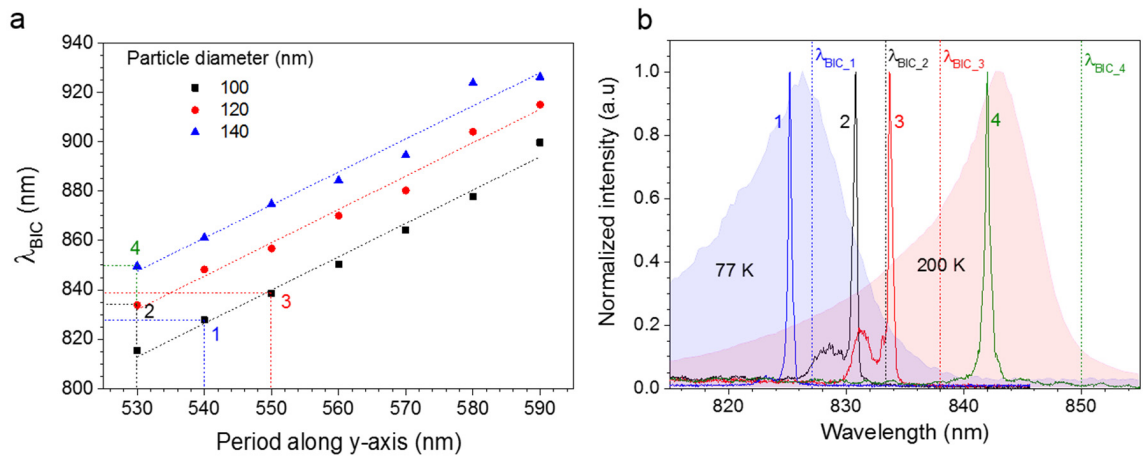

**Figure 4. Wavelength tunable lasing in dielectric nanoantenna array. a**, Resonance wavelength at quasi-BIC condition for arrays with different geometrical parameters (nanopillar diameter and array period along y-axis). The colored numbers are corresponding to the arrays chosen for lasing measurement in b. **b**, Lasing spectra for different arrays 1-4 with parameters as indicated in a. The lasing measurements for arrays 1-3 were performed at 77 K while for array 4, it was performed at 200 K. This is done to match the gain spectrum of GaAs with designed resonance wavelength of the selected arrays, as shown by the photoluminescence spectra of GaAs for different temperatures plotted as shaded areas.

One of the advantages of using the vertical dipole resonance mode in our laser structure is that the leaky resonance from the BIC condition can be satisfied with a wide range of geometrical parameters. By tuning either the particle diameter or the y-axis period, the quasi-BIC resonant wavelength can be precisely tuned. **Figure 4a** shows the leaky resonance wavelength for various arrays with different particle sizes and periods, extracted from angle resolved transmission measurement (see Supplementary Information, Figure S10 for full data). Also by changing the temperature from 77 K to 200 K, the peak of the gain spectrum of GaAs can be shifted from 830 nm to 850 nm (Supplementary Information, Figure S11). Combining these two effects it is possible to achieve lasing at different selected angles and wavelengths from different resonant arrays. **Figure 4b** shows lasing spectra achieved for 4 different arrays (corresponding to the same color-dashed circles in Figure 4a) at 77K (for arrays 1-3), and 200 K (for array 4). Attempts to achieve lasing at higher temperature were not successful because of the low gain of GaAs caused by its high surface recombination. We believe that by using higher gain material or improving the emission quantum efficiency of GaAs (*e.g.,* by surface passivation) it should become possible to achieve lasing at room temperature.

In conclusion, we have demonstrated directional lasing action in arrays of active dielectric nanoantennas for the first time. This is achieved using vertical electric dipole resonances excited inside GaAs nanopillars, which, in the purely non-diffractive case, would form a non-radiative bound state in the continuum. The directionality of the laser is controlled by a leaky channel, opened by designing one period of the array to support diffractive orders. The lasing happens at the angle where the emission bands (defined by the leaky channel) present a crossing point in the emission plane. Despite the presence of the radiation channel, sufficiently high Q factors can be retained for small lasing angles (*i.e.*, <10°). Moreover, lasing can also be obtained at an angle as large as 25 degrees to the normal. This is achieved by temperature-induced tuning of the gain spectrum to compensate for the lower Q factor corresponding to a larger angle and leads to a minimum lasing threshold at the corresponding wavelength. By tuning geometrical parameters of the array such as particle size or period, the leaky resonance wavelength can be precisely controlled to achieve lasing at different wavelengths. This design concept can be applied to other high-index active semiconductor materials and be readily integrated in multi-layered photonic devices due to its high transparency. Our results demonstrate that Mie resonances in dielectric nanoparticles can be used to achieve stimulated emission even in relatively low gain materials such as un-passivated GaAs. This brings new opportunities to nanophotonic research by providing a novel platform for making highly-efficient directional emitting devices based on active dielectric nanoantennas.

## Author Contributions

S.T.H. and Y.H.F. constructed the low-temperature emission measurement setup and performed the sample characterization. S.T.H. wrote the first draft of the manuscript. N.K.E. optimized the fabrication process to prepare the sample, and performed initial stage optical characterization and analysis. Z.P. assisted in constructing the measurement setup. R.B. helped on initial stage optical measurements. R.P.D. conceived the idea of bound state in the continuum laser in the nanoantenna arrays and performed the simulations. A.I.K. conceived the idea of dielectric nanoantenna lasers and coordinated the work. All authors discussed the results and worked on the manuscript.

## Funding Sources

A*STAR SERC Pharos program, Grant No. 152 73 00025 (Singapore).

## Acknowledgements

We acknowledge Vytautas Valuckas (DSI) for SEM characterization.

# Supplementary Information:

# Lasing action in active dielectric nanoantenna arrays

*Son Tung Ha[†], Yuan Hsing Fu[†], Naresh Kumar Emani[†§], Zhenying Pan, Reuben M. Bakker, Ramón Paniagua-Domínguez, and Arseniy I. Kuznetsov\**

Data Storage Institute, A*STAR (Agency for Science, Technology and Research), 138634, Singapore

*Corresponding author, email: <u>arseniy_k@dsi.a-star.edu.sg</u>

[†]These authors equally contributed to this work.

[§]Present affiliation: Indian Institute of Technology, Hyderabad, India

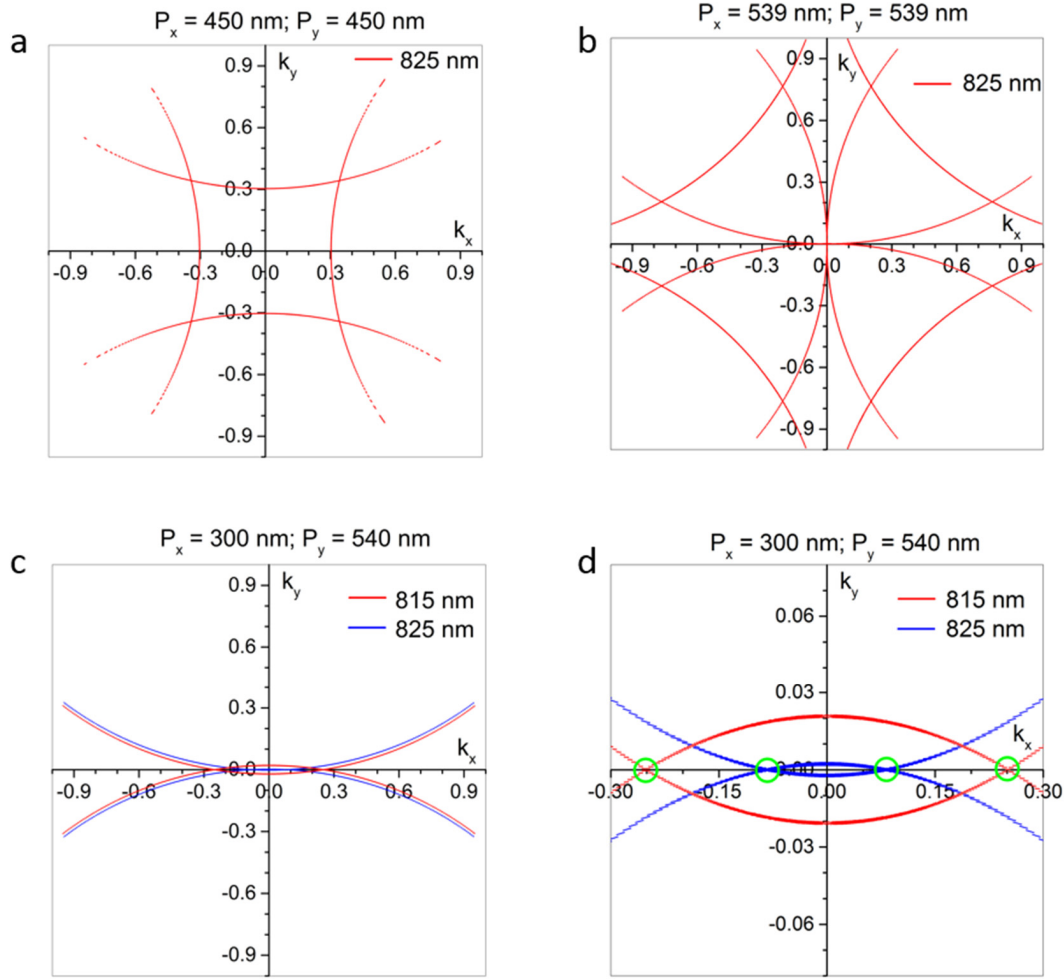

**Figure S1. Theoretical calculation of the emission bands for a 2D GaAs array with different periods and wavelengths based on reciprocity.**[1] a, For Px = 450 nm, Py = 450 nm and 825 nm wavelength (subdiffractive at normal incidence/emission). b, For Px = 539 nm, Py = 539 nm and 825 nm wavelength (supporting diffractive modes at normal incidence/emisison). c, For Px = 300 nm, Py = 540 nm and 815 nm (red) and 825 nm (blue) wavelengths (below diffraction in x-axis and slightly above diffraction in y-aixs). d, Zoom-in emission band in c, showing the emission cross-point at $k_x$ at small angle for 825 nm emission and larger angle for 815 nm emission.

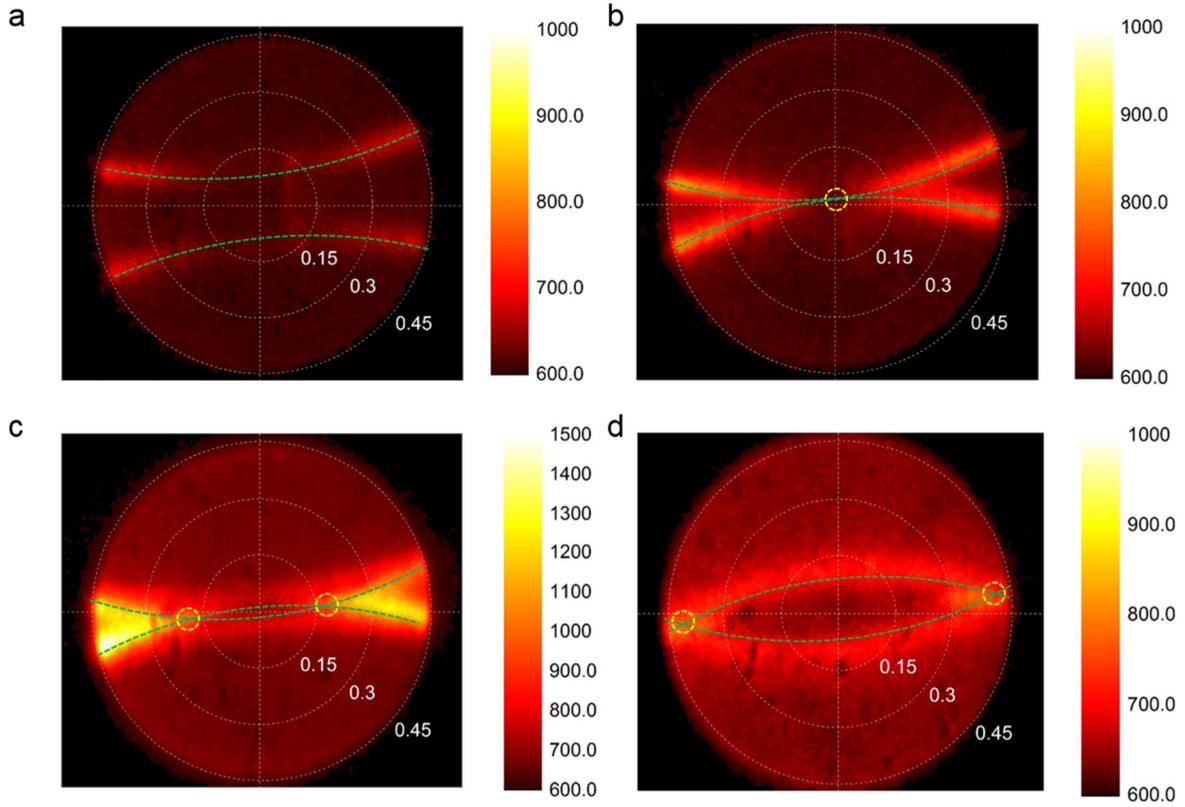

**Figure S2. Experimental back focal plane emission patterns**. a, $P_y$ = 530 nm (subdiffractive at normal incidence/emission). b, $P_y$ = 540 nm (supporting diffraction at normal incidence/emisison). c, $P_y$ = 550 nm – (slightly above diffraction at normal incidence/emission). d, $P_y$ = 590 nm – (above diffraction at normal incidence/emission). All results are for emission wavelength around 830 nm. In all cases the period along the x-axis, $P_x$, is fixed at 300 nm. The signals were collected using long working distance objective (50X, WD = 17 mm) with numerical aperture NA = 0.45 resulting in a maximum collection angle of ~ 28 degree.

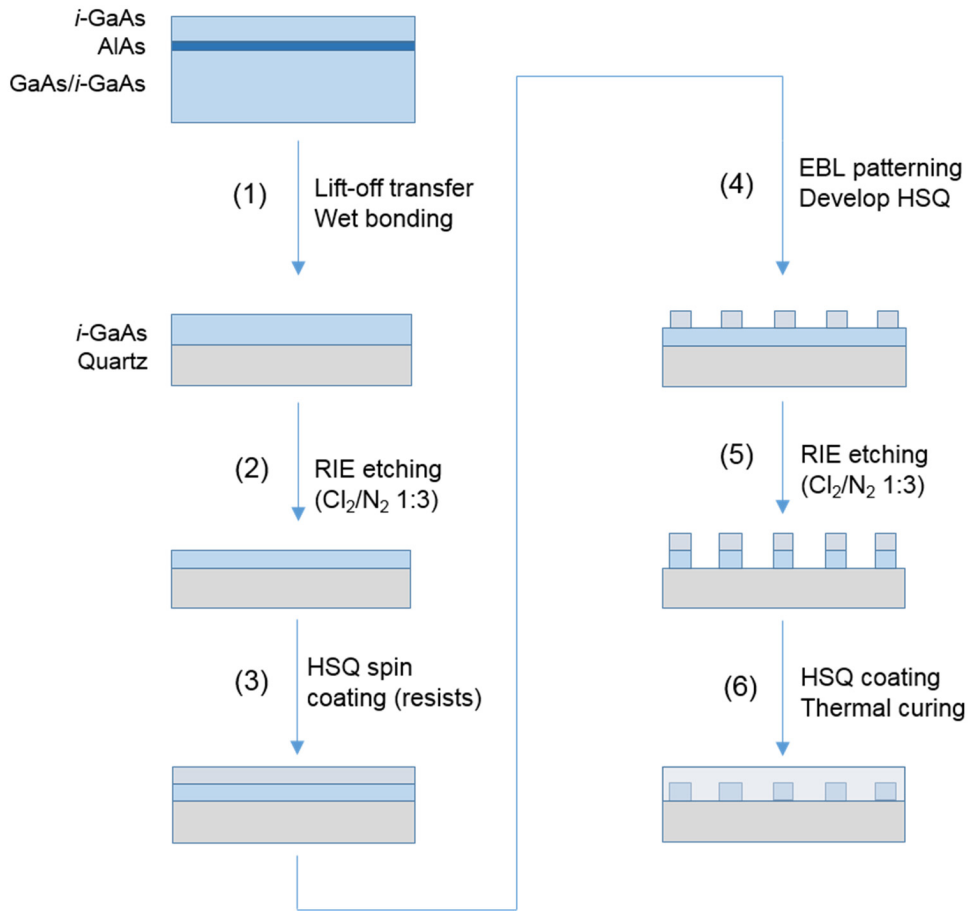

**Figure S3. Fabrication steps for GaAs nanopillar arrays.** (1) The fabrication starts with a *i*-GaAs/AlAs/*i*-GaAs wafer purchased from Semiconductor Wafer, Inc (Taiwan). The bottom GaAs (substrate) layer consists of 625 μm undoped <100> GaAs. The middle layer is AlAs with a thickness of 100 nm. The top layer is *i*-GaAs with a thickness of 500 nm. To obtain the GaAs film we follow the epitaxial lift-off process described by Yablonovitch *et al*.[2,3] In the process black wax (*i.e.,* Apiezon W) is applied on the top *i*-GaAs layer and thermally annealed at 90°C for about 30-60 mins to realize a slightly domed surface. The sample is then placed in dilute hydrofluoric acid (HF) solution (~ 5 wt. % in $H_2O$) overnight to selectively remove the AlAs layer. The top *i*-GaAs layer along with the wax detaches from the wafer and floats in the aqueous etchant solution. The *i*-GaAs film is then gently placed on a fused silica substrate (quartz) and the interfacial water layer is left to dry out. When the water is evaporated the GaAs film will stick to underlying quartz due to Van der Waals forces. Finally, the wax layer is removed by placing the sample in Dichloromethane (other polar solvents like Xylene and Trichloroethylene will work as well) and cleaned Acetone, IPA and $H_2O$. (2)

Inductively Coupled Reactive ion etching (ICP-RIE, Oxford Plasmapro) using $Cl_2$ chemistry is used to etch the initial thickness of *i*-GaAs (*i.e.,* 500 nm) to about 250 nm as in our design. The GaAs film thickness before and after etch was monitored using spectral reflectance measurements (Filmetrics F20). (3) A negative resist – Hydrogen silsesquioxane (HSQ, XR1541-6, Dow Corning) is coated directly onto *i*-GaAs at 5000 RPM for 90 seconds and baked on a hotplate at 140°C for 3 minutes. (4) Electron beam lithography is used to pattern the nanopillar array into HSQ. The HSQ film is then developed in 25% Tetramethyl Ammonium Hydroxide(TMAH) in $H_2O$. (5) ICP-RIE is used again to obtain GaAs pillar pattern using the patterned HSQ as the mask. (6) The GaAs arrays were covered in a thick layer of HSQ resist by repeated spin-coating and thermal curing steps. The curing process cross links and hardens the HSQ resist resulting in the formation of Spin On Glass with refractive index close to 1.5.

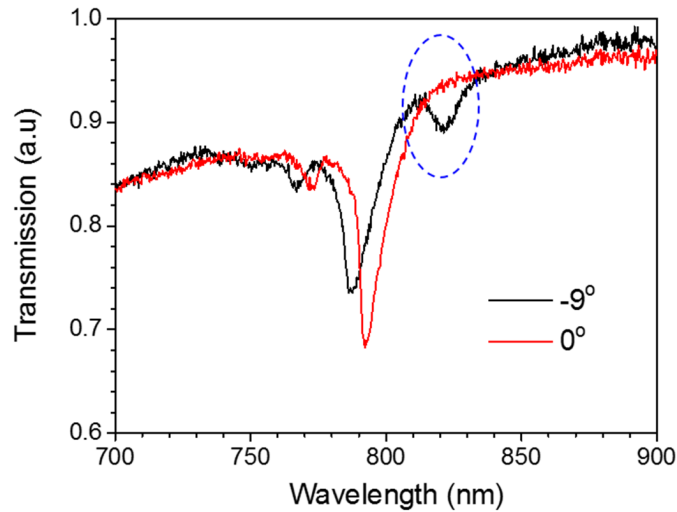

**Figure S4. Transmission spectra at different angle.** Transmission data extracted from Figure 2b in the main text for 2 different angles: 0° and -9°. As can be seen, the resonance at ~ 825 nm vanishes at 0 degree (blue-dash circle) while the other mode at ~ 790 nm is just red-shifted for decreasing angle of incidence.

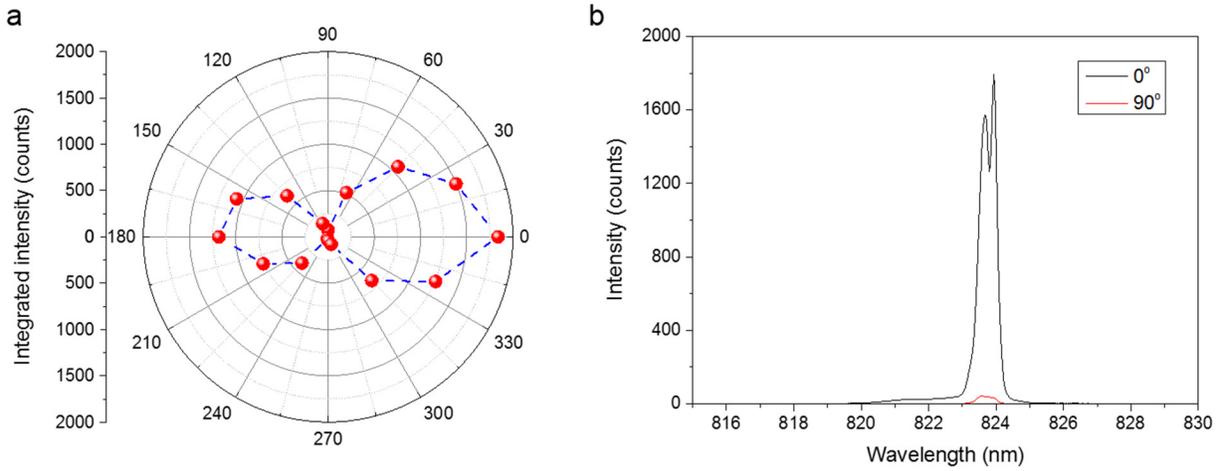

**Figure S5. Polarization dependence of the lasing.** The lasing signal after the collection objective is passed through a polarizer with controlled polarization angle before being collected by the spectrometer. We defined 0° (180°) angle to correspond to direction along y axis of the array. a, Lasing intensity dependence on polarization angle showing maximum at 0 and 180° (along x axis), consistent with the vertical dipole orientation, from which the leaky resonance emerges, and the lasing directivity as discussed in Figure 3 in the main text. The asymmetry of the polarization dependence may come from variations in the signal collection and/or pumping laser power fluctuations. b, Lasing spectra for polarization at 0 and 90° showing at least 2 orders of magnitude difference in the intensity.

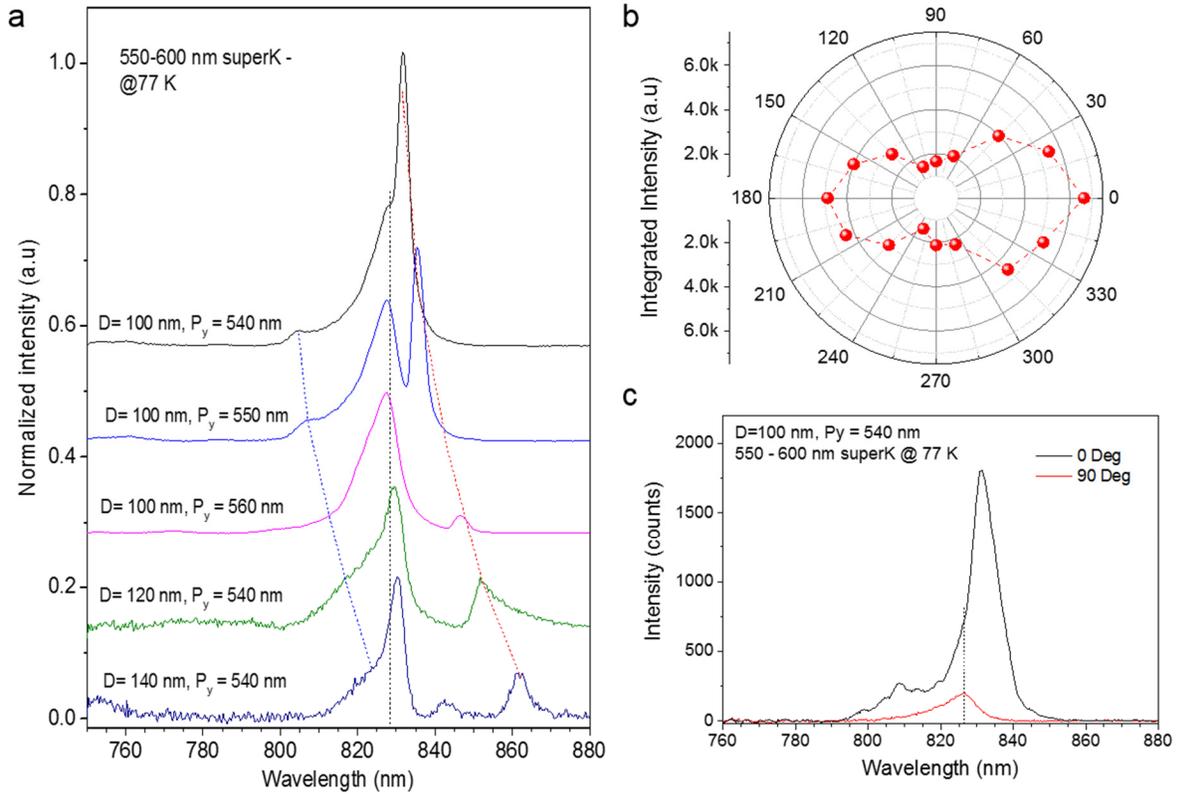

**Figure S6. Resonantly enhanced photoluminescence study.** In this experiment we used a supercontinuum laser (superK, NKT) as a pumping source. The pumping wavelength range of 550- 600 nm was selected using a variable bandpass filter. This laser has 80 MHz repetition rate and a pulse duration of a few hundred picoseconds. This results in a much lower laser fluence compared to the femtosecond laser source used in lasing experiments. Nevertheless, resonantly enhanced photoluminescence peaks can be observed. **a,** Emission spectra measured for different GaAs nanopillar arrays with parameters labeled in the figure (D – diameter of the nanopillar; $P_y$ – array period in y direction; $P_x$ and height of the nanopillars were kept at 300nm and 250 nm, similar to the lasing array studied in the main text). They show a photoluminescence peak at around 828 nm and 2 sets of narrower peaks enhanced by the horizontal dipole resonance (shorter wavelength peak) and the vertical dipole resonance (longer wavelength peak) which is consistent with the angle resolved transmission measurement in Figure 2b in the main text. For the lasing array with $D$ = 100nm and $P_y$ = 540 nm the vertical dipole-enhanced peak is overlapped with the PL peak of GaAs (at 77K). This explains why the emission band for the lasing array vanishes near 0°, as shown in Figure 3c in the main text, as the photoluminescence is resonantly enhanced by vertical dipole

resonance, which does not emit in the vertical direction. **b** and **c,** polarization dependence of resonant enhanced photoluminescence showing similar behavior as the lasing emission shown in Figure S4.

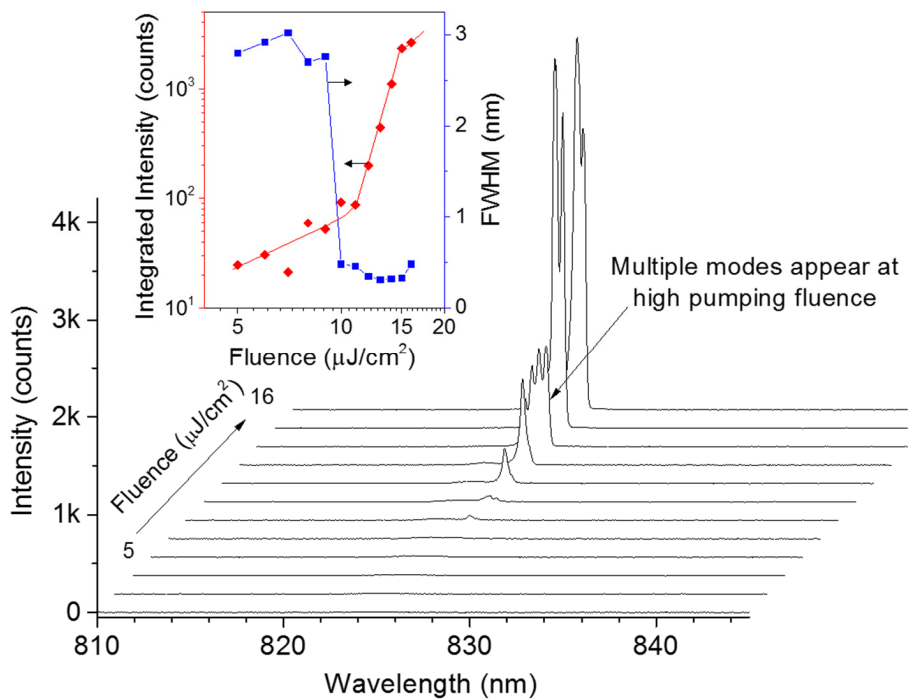

**Figure S7. Multiple mode lasing at higher pumping fluences.** When pumping at high fluence (*i.e.*, > 14 µJ/cm$^2$ in the sample where the lasing threshold is ~10 µJ/cm$^2$), multiple mode lasing is observed. At 14 µJ/cm$^2$, up to 4 lasing modes are observed while at even higher pumping fluence, only the modes at longer wavelengths survive. Inset: Integrated output intensity as a function of input fluence in log-log scale and full width at half maximum (FWHM) of the emission peak.

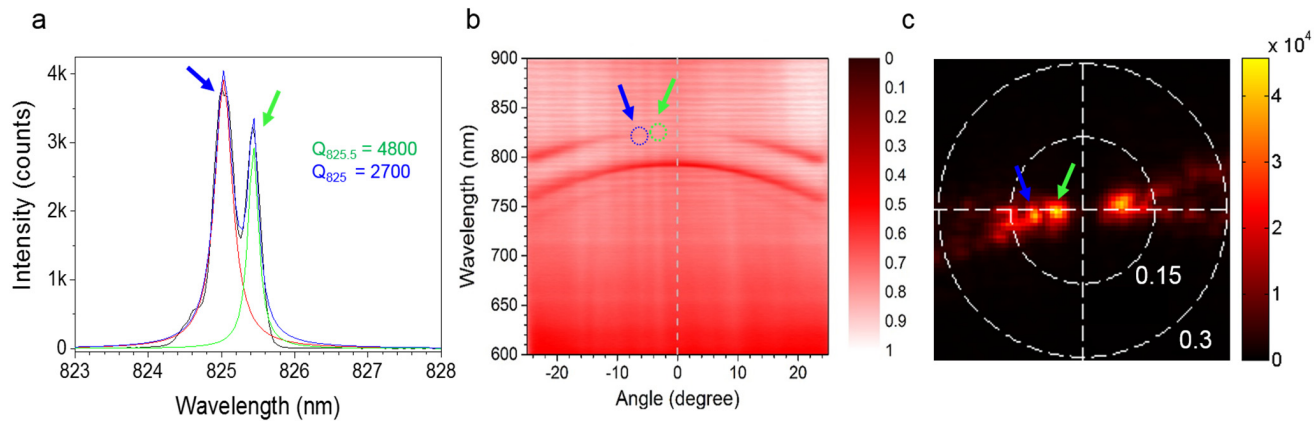

**Figure S8. Analysis of higher mode lasing at high pumping fluence**. **a**, Fitting of lasing spectrum at 15 µJ/cm² pumping fluence reveals two lasing modes with different FWHM: Peak 1 at 825.03 nm has Q = 2700 (marked by a blue arrow) while peak 2 at 825.43 nm has Q = 4800 (marked by a green arrow). **b**, Angle resolved transmission spectrum of the sample showing the multi-mode lasing action positions corresponding to panel a. At higher pumping fluence, an additional resonant mode, corresponding to a higher emission angle and shorter wavelength in the resonance band can be co-excited. The primary and secondary modes are depicted with blue and green dashed circles in the angle-resolved transmission spectrum. **c**, Back focal plane image of the directionality of the multi-mode lasing, confirming the emergence of a secondary emission angle, indicated by a blue arrow, corresponding to the lasing mode at shorter wavelength (*i.e.,* 825.03 nm). The angle of this mode, as expected, is larger than that of the mode at longer wavelength (depicted as a green arrow).

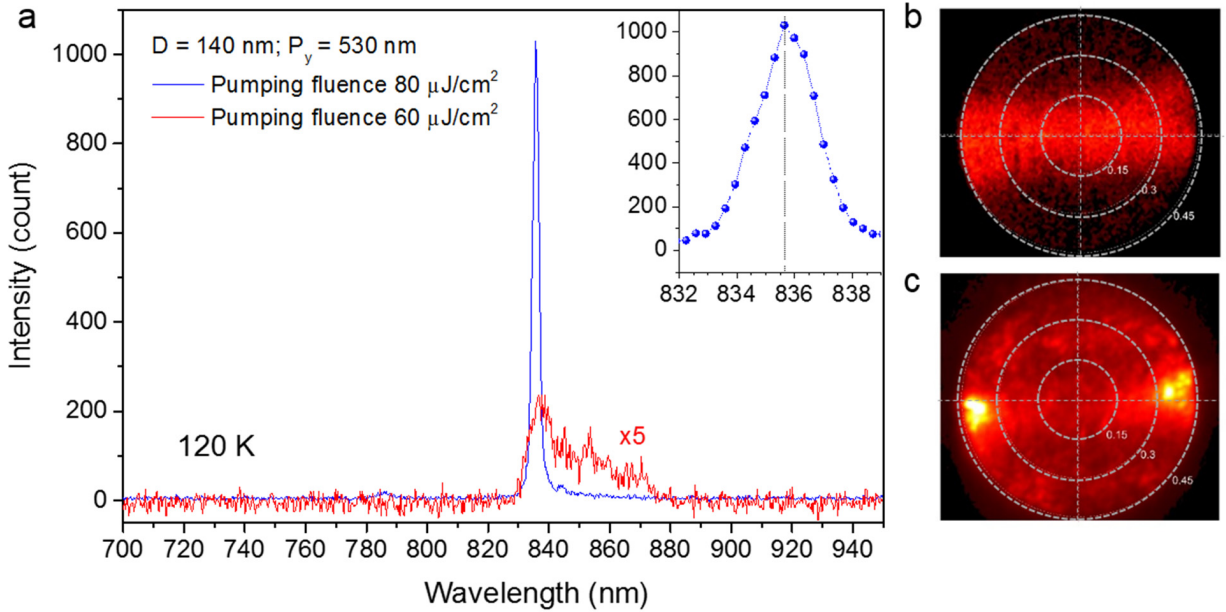

**Figure S9. Lasing at higher angle in a 2D GaAs array.** In these measurements a GaAs array with the following parameters is used: $D$ = 140 nm and $P_y$ = 530 nm which has $\lambda_{BIC}$ = 850 nm, as shown in Figure 4a in the main text. By changing the temperature to 120 K, for which the PL of GaAs is peaked at 835 nm, we were able to achieve lasing at ~ 836 nm with a lasing angle of ~ 26 degree. **a**, Emission spectra of the GaAs array before (red) and after (blue) the lasing threshold when pumped with a 780 nm femtosecond laser (20 KHz repetition rate, 200 fs pulse width). For lasing experiment at higher temperature than 77 K, we employed a lower repetition rate (*i.e.,* 20 KHz vs 100 KHz at 77 K) to minimize the accumulation of heat leading to sample damage. Inset: zoomed-in lasing peak showing the FWHM of around 2 nm, corresponding to a Q factor of 410. **b** and **c**, Back focal plane images of emission below (b) and above (c) lasing threshold. Above the lasing threshold, two enhanced emission points in the emission plane are clearly seen, corresponding to a lasing angle of around 26 degrees. As discussed in the main text, the lasing happens at the wavelength and angle where a minimum lasing threshold $P_{th}(\lambda) \sim (g(\lambda) \times \Gamma_E(\lambda))^{-1}$ is achieved. In our system, it is possible to tune the gain spectrum of GaAs by changing the device temperature to achieve the minimum threshold, and thus lasing, at a higher angle, as exemplified here.

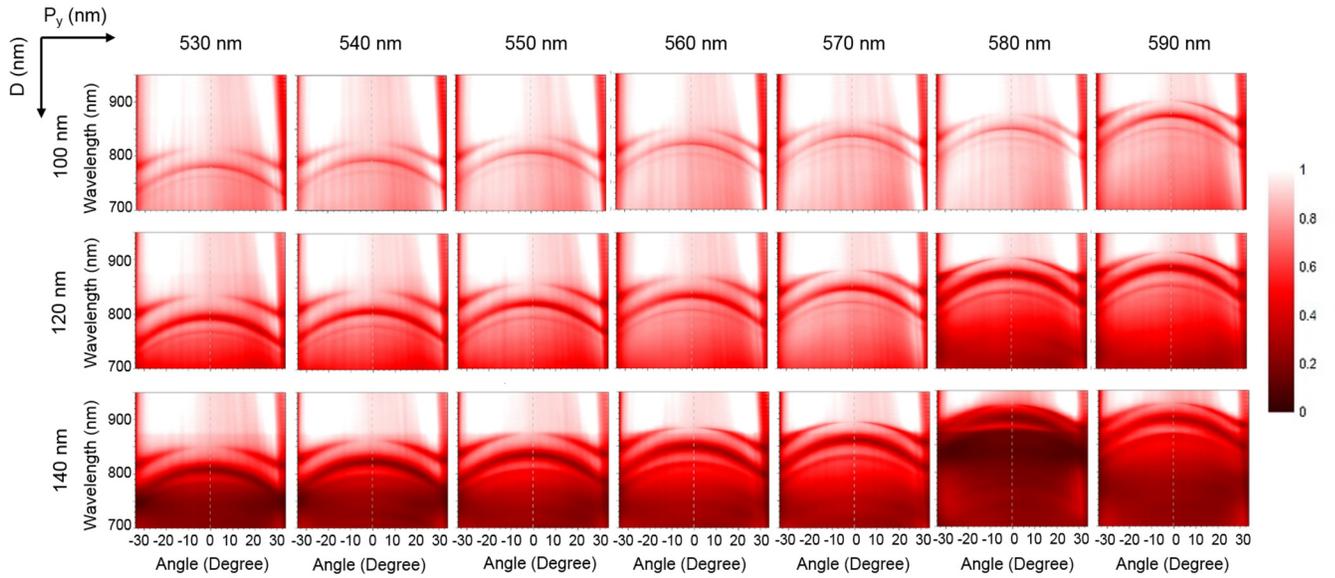

**Figure S10. Angle resolved transmission measurements for different diffractive array periods and nanopillar diameters (specified in the plot).** The geometry of measurements is the same as discussed in Fig.2 in the main text. $\lambda_{BIC}$ is determined by fitting the transmission dip corresponding to the vertical dipole resonance (*i.e.,* the longest wavelength dip) at 0° angle. The determined $\lambda_{BIC}$ of the arrays with different periods and particle sizes are summarized in Figure 4a in the main text.

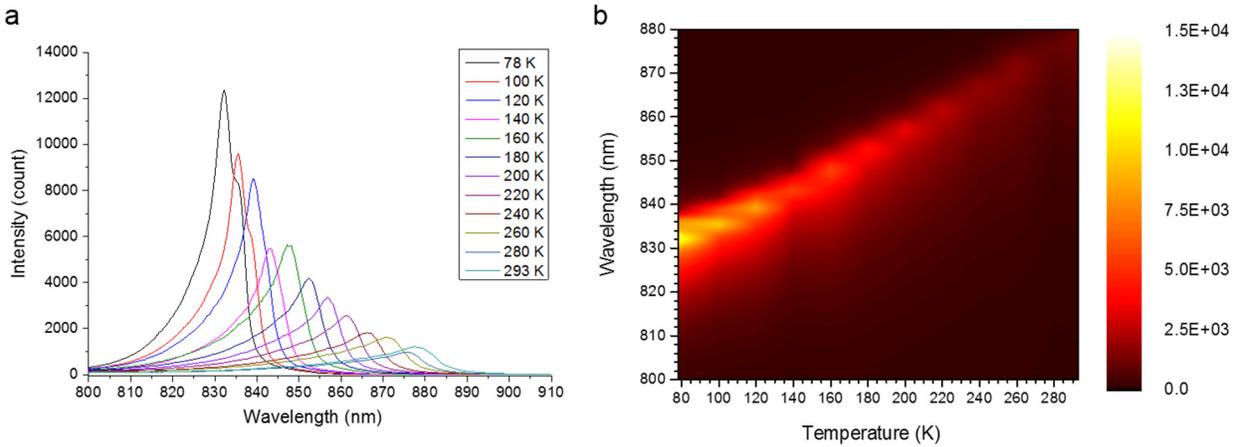

**Figure S11. Temperature dependent photoluminescence of GaAs films**. **a,** PL spectra evolution of a GaAs film at different temperatures. b, Contour plot of the temperature dependent PL (same as in a) is represented to show the emission peak shift and increase of gain (*i.e.*, of the PL intensity) as the temperature decreases. The PL spectra shown correspond to GaAs films. The PL spectra for nano-sized GaAs pillars are blue-shifted for around 5-10 nm compared to the films, as shown in Figure S5 for the case of PL at 77K. In this case the PL peak of GaAs films happens at ~ 832 nm while for the nanopillar array the PL is peaked at ~ 827 nm. This red-shift may be originated from the heating effect of the GaAs film during laser excitation.

**Numerical simulations**

Numerical simulations presented in the paper were performed using a commercial solver based on the Finite-Element Method (FEM), COMSOL Multiphysics©. The simulation domain consists on a rectangular box, representing a unit-cell of the array, and a cylinder representing the GaAs nanopillar. Periodic Bloch boundary conditions are applied in the boundaries along the x- and y-directions to mimic an infinite, two-dimensional system. The periodicities of the array are $P_x$ = 300 nm and $P_y$ = 540 nm (corresponding to the physical size of the rectangular box in the x- and y-directions, respectively). Port boundary conditions are used in the top and bottom boundaries to excite the system and to collect outgoing waves both in transmission and reflection. As many ports as diffraction orders supported by the lattice are used both in transmission and reflection. The excitation field corresponds to that of a p-polarized plane wave incident onto the array at a certain angle, $\theta$, contained in the xz-plane. The total size of the rectangular box is 1 µm in the z-direction. GaAs particles are modeled as cylinders with height $H$ = 250 nm and diameter $D$ = 100 nm and are considered to be embedded in a background material with refractive index of 1.45. The material parameters used for GaAs are those experimentally measured by ellipsometry in the GaAs films used. In the transmission results presented in Figure 2d of the main text only the direct transmission is plotted.

The multipole decomposition presented in Figure 2e in the main text is based on the decomposition of the particle internal fields using a Cartesian basis. The electric dipole moment excited in the particle can be calculated as:

$$\mathbf{p} = \int \varepsilon_0 (\varepsilon - \varepsilon_d) \mathbf{E} d\mathbf{r}$$

In this expression, $\varepsilon_0$ is the permittivity of vacuum, $\varepsilon_d$ and $\varepsilon$ the relative permittivities of the surrounding medium and the particle, respectively, $\mathbf{E} = \mathbf{E}(\mathbf{r})$ the vector electric field and the integral is taken over the volume of the particle. From this calculation one can readily identify the main component of the electric dipole induced in the particles, as done in Figure 2e, in which both horizontal and vertical induced electric dipoles could be identified. From the dipole moments it is then possible to compute the scattering cross section associated with them as:

$$C_{ED} = \frac{k_0^4}{6\pi \varepsilon_0^2 E_0^2} |\mathbf{p}|^2$$

The rest of the multipole moments (not shown in the paper for being negligible) can be computed as well from the internal fields of the particle following the expressions detailed elsewhere.[4]